\documentclass[%
 aip,jap,
 amsmath,amssymb,
preprint,%
]{revtex4-1}

\usepackage{graphicx}
\usepackage{dcolumn}
\usepackage{bm}
\usepackage[english]{babel}
\usepackage{epstopdf}

\usepackage[latin1]{inputenc}
\usepackage[T1]{fontenc}
\usepackage{color}

\begin{document}

\preprint{AIP/123-QED}

\title{Determination of the Fermi Level Position \\ in Dilute Magnetic Ga$_{1-x}$Mn$_{x}$N Films}

\author{S. Barthel}
\email[]{sbarthel@itp.uni-bremen.de}
\affiliation{Institute for Theoretical Physics, University of Bremen, D-28359 Bremen, Germany}

\author{G. Kunert}
\affiliation{Institute of Solid State Physics, University of Bremen, D-28359 Bremen, Germany}

\author{M. Gartner}
\author{M. Stoica}
\affiliation{Institute of Physical Chemistry ``Ilie Murgulescu'', Romanian Academy, 060021 Bucharest, Romania}

\author{D. Mourad}
\affiliation{Institute for Theoretical Physics, University of Bremen, D-28359 Bremen, Germany}

\author{C. Kruse}
\affiliation{Institute of Solid State Physics, University of Bremen, D-28359 Bremen, Germany}
\affiliation{Department of Physics, University of Osnabrück, D-49076 Osnabrück, Germany}

\author{S. Figge}
\affiliation{Institute of Solid State Physics, University of Bremen, D-28359 Bremen, Germany}

\author{D. Hommel}
\affiliation{Institute of Solid State Physics, University of Bremen, D-28359 Bremen, Germany}

\author{G. Czycholl}
\affiliation{Institute for Theoretical Physics, University of Bremen, D-28359 Bremen, Germany}

\date{\today}

\begin{abstract}
We report on a combined theoretical and experimental determination of the Fermi level position in wurtzite Ga$_{1-x}$Mn$_{x}$N films with $x=4\%$ and $x=10\%$ as grown by molecular beam epitaxy. By means of ellipsometric measurements the real part of the frequency-dependent conductivity is determined. An electronic model in the framework of the effective bond-orbital model is parameterized in order to theoretically reproduce the measured transport properties. Predictions for the long-wavelength behaviour as a function of the Fermi level are made. The corresponding density of states obtained in this model is in qualitative agreement with first-principle calculations. The absence of a significant experimental peak in the AC conductivity for small frequencies indicates that the Fermi level lies in a gap between two Mn-related impurity bands in the host band gap.
\end{abstract}

\maketitle


\section{Introduction}
According to early theoretical work\cite{dietl_zener_2000}, the dilute magnetic semiconductor (DMS) Ga$_{1-x}$Mn$_{x}$N is supposed to be a ferromagnetic material with a high Curie temperature $T_C$. The question whether or not this description is correct has been adressed both from an experimental as well theoretical point of view by various authors\cite{bonanni_ferromagnetic_2007,sato_first-principles_2010,bouzerar_unified_2010,chakraborty_nanoscale_2012} and it appears that ferromagnetic order above room temperature does not occur in samples without a large amount of impurity clusters: Recent reports\cite{bonanni_experimental_2011,sawicki_origin_2012} about Ga$_{1-x}$Mn$_{x}$N layers deposited by metal-organic vapor-phase epitaxy (MOVPE) with Mn concentrations $x$ ranging from 0.5\% up to 3.1\% indicate a homogeneous incorporation of impurities on Ga-sites in combination with an electronic configuration of $3d^{4}$ for the Mn ions. The reported critical temperature is low with $T_C\approx1.8$ K for $x\approx3.1\%$ and the mechanism leading to collective magnetic order was attributed to ferromagnetic superexchange while the samples also are highly resistive. On the other hand Kunert \textit{et al.}\cite{kunert_ga1xmnxn_2012} observed large magnetizations and Curie temperatures of $T_C\le14$ K for wurtzite Ga$_{1-x}$Mn$_{x}$N layers produced by molecular beam epitaxy (MBE) with Mn concentrations up to $x=10\%$ and most likely the spin $S=2$ configuration\cite{sawicki_origin_2012} for homogeneously incorporated Mn impurities. The knowledge about the electronic structure and position of the Fermi level is of major interest in order to determine the possibly insulating/half-metallic nature of the Ga$_{1-x}$Mn$_{x}$N material with high Mn content.
Furthermore, this knowledge is useful for codoping experiments with Mg or Si in order to establish an exchange mechanism based on direct or $sp/d$-exchange enabling high Curie temperatures.
Though first principle approaches are in general the tool of choice for electronic structure calculations, the inclusion of electron correlations within the open shell of the transition metal ions in DMS can easily lead to different results: For instance, Virot \textit{et al.}\cite{virot_electronic_2011} report on insulating behaviour for GaN:Mn within local spin-density approximation and Coulomb interaction (LSDA+U), whereas LSDA alone leads to a semimetal. In addition, \textit{ab initio} approaches are limited by the supercell size for the simulation of disorder effects in real-space.

In this work an empirical tight-binding approach based on the effective-bond-orbital model (EBOM) of the III-V wurtzite nitride material system\cite{mourad_multiband_2010} is chosen in order to calculate the electronic and transport properties of the DMS Ga$_{1-x}$Mn$_{x}$N.

Thereby, the difficulties in first principle calculations regarding the role of correlations can be avoided and the lattice model can account for disorder effects by means of an ensemble average over different configurations with random impurity positions.
For the simulation of magnetic impurities, the EBOM is augmented by a potential scattering and a spin-spin interaction term (V-J model)\cite{bouzerar_non-perturbative_2007,barthel_effective_2013}. Then the frequency-dependent conductivity is evaluated via the kernel-polynomial method (KPM)\cite{weise_kernel_2006} in linear response theory as a function of the Fermi level position. A direct comparison to results based on ellipsometric measurements on samples fabricated by Kunert \textit{et al.}\cite{kunert_ga1xmnxn_2012} enable conclusions about the density of states (DOS) and Fermi-Level position of Ga$_{1-x}$Mn$_{x}$N.\\

The paper is structured as follows: First, the experimental setup for the ellipsometric characterization is explained, followed by a description of the applied theoretical approaches. Then the experimentally determined optical constants between 200 nm and 12400 nm of Ga$_{1-x}$Mn$_{x}$N layers with $x=4\%$ and $x=10\%$ and corresponding theoretical results for $x=10\%$ of the real part of the AC-conductivity, DOS and Fermi level are discussed. The paper concludes with a summary of the main results.

\section{Experiment}

\subsection{Ellipsometric setup}
Spectroscopic ellipsometer (SE) measurements were performed in order to obtain the thickness, real and imaginary part of the complex refraction index ($n,\kappa$) on a broad spectral range, using \textit{Woollam Co., Inc.} equipment composed of a rotating-analyzer VASE ellipsometer for UV-VIS-NIR range and a rotating-compensator infrared spectroscopic ellipsometer for IR spectral range (IR-SE). Measurements have been performed at room temperature, using the 60$^\circ$ incidence angle for both spectral ranges, 2 nm wavelength step in UV-VIS-NIR range and 8 cm$^{-1}$ resolution for IR range (2-12.4 $\mu$m).\\

The ellipsometric measurements were fitted with a four layer model: sapphire substrate / GaN buffer layer / Ga$_{1-x}$Mn$_{x}$N film / roughness layer, where the optical constants of sapphire-c-cut biaxial were taken from the \textit{Woollam} data base, the GaN template used in the computations were previously measured on our reference samples. For Ga$_{1-x}$Mn$_{x}$N layers the general oscillator model (containing Gauss, TOLO, rho-tau and PSMO oscillators\cite{tompkins_wvase32_2006}) in combination with the Simple Gradient Index model was used. The roughness layer was described with the Effective Medium Approximation Model\cite{bruggemann_berechnung_1935} considering 50\% voids and 50\% GaMnN. The best match between the experimental and modeled data was achieved through a least-square-regression analysis.
From the best fit, the thickness and optical constants of Ga$_{1-x}$Mn$_{x}$N layers were determined. The real part of the frequency dependent conductivity with $\omega$ as the photon energy is then given by (we set $\hbar=1$ throughout this paper):
\begin{equation}
\sigma_{\text{exp}}^{\text{AC}}(\omega)=\frac{\omega}{2\pi}n(\omega)\kappa(\omega).
\end{equation}

\subsection{Results}
\begin{figure}
\includegraphics[width=\columnwidth]{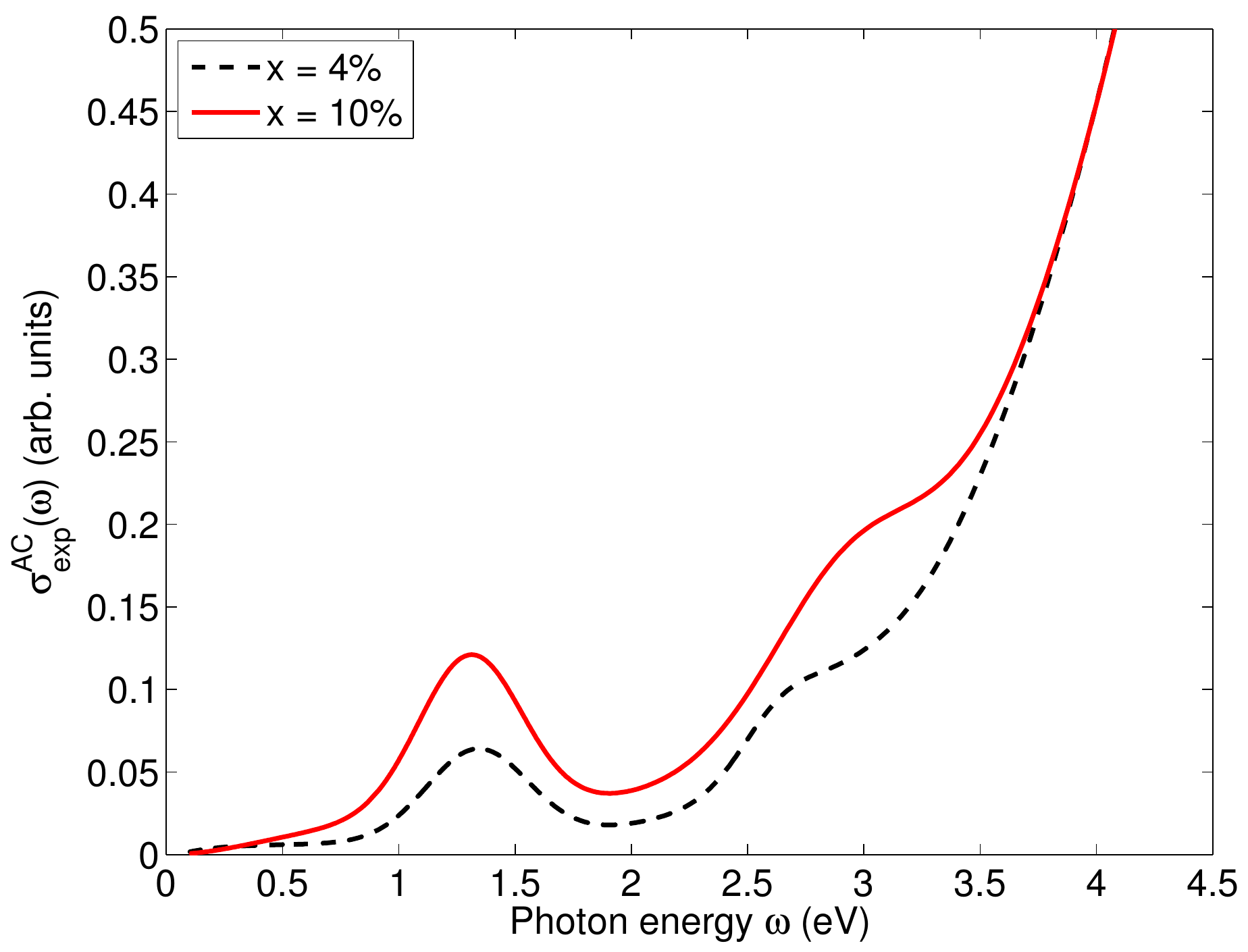}
\caption{\label{fig_1}Experimentally determined real part of the AC-conductivity of Ga$_{1-x}$Mn$_{x}$N for several Mn-concentrations $x$ based on ellipsometric data taken at room temperature.}
\end{figure}
The experimental results for the real part of the frequency-dependent conductivity $\sigma_{\text{exp}}^{\text{AC}}(\omega)$ are shown in Fig.\ref{fig_1} for the two Mn-concentrations $x=4\%$ and $x=10\%$. Three characteristic features for photon energies $\omega>0.8$ eV can be observed for both concentrations: a) A steep slope is present from high energies of $\omega >4$ eV to lower energies, crossing the energy axis when extrapolated slightly above 3 eV in qualitative agreement with an expected host GaN band gap of $\sim3.4$ eV\cite{maruska_preparation_2003} at room temperature. b) A small shoulder is located in the vicinity of 2.9 eV falling off at about $\sim 2$ eV and c) a pronounced peak at $\omega=1.31$ eV ($x=10\%$) and $\omega=1.35$ eV ($x=4\%$). The features observed in our measurements are characteristic for Ga$_{1-x}$Mn$_{x}$N\cite{graf_mn3+/2+_2002,korotkov_optical_2002,graf_growth_2003,graf_prospects_2003}. Graf \textit{et al.} attribute the cases b) and c) to excited states resulting from different ionization processes of the Mn-impurities. b) corresponds to Mn-impurities with a Mn$^{3+}$-configuration with a transition to Mn$^{2+}+\text{hole}$ via direct photoionization. c) is interpreted as an internal optical excitation from Mn$^{3+}$ in (Mn$^{3+}$)$^{*}$ with an energy of about 1.42 eV and followed up by thermal ionization to Mn$^{2+}+\text{hole}$ via an activation energy of $\approx0.3$ eV.
Very recent results from absorption measurements, measured in transmission, also point to a comparable absorption profile, including a local maximum at $\sim 1.4$ eV and a shoulder with its onset at $\sim 2$ eV (not shown).\cite{papierska_private_2014}
\section{Theory}

\subsection{Electronic Model}
In order to model the electronic properties of the dilute magnetic semiconductor Ga$_{1-x}$Mn$_{x}$N with $x=10\%$ the following multiband V-J model Hamiltonian\cite{bouzerar_non-perturbative_2007,barthel_effective_2013} is used:
\begin{eqnarray}
\hat{H}&=&\sum_{ij,\alpha\beta,\sigma}t_{ij}^{\alpha\beta,\sigma}\hat{c}_{i\alpha\sigma}^{\dagger}\hat{c}_{j\beta\sigma}\nonumber\\
&+&\sum_{i,\alpha,\sigma}p_{i}V_{\alpha\alpha}\hat{n}_{i\alpha\sigma}-\sum_{i,\alpha}p_{i}J_{\alpha d}\hat{\mathbf{S}}_{i}\hat{\mathbf{s}}_{i,\alpha}.
\label{eq:H_gamnn}
\end{eqnarray}
Its first term corresponds to the tight-binding description of the electronic properties of the host material GaN with the hopping matrix-elements $t_{ij}^{\alpha\beta,\sigma}$ between different lattice sites $i,j$, orbitals $\alpha,\beta$ and spin $\sigma$, but the effect of spin-orbit coupling is neglected, as the spin-orbit splitting is small\cite{vurgaftman_band_2003} ($\Delta_{\mathrm{so}}=0.017$ eV) in the material under consideration. As we specifically employ the parametrization from Ref.\onlinecite{mourad_multiband_2010} of the wurtzite III-V nitrides for the $t_{ij}^{\alpha\beta,\sigma}$, a minimal $sp^3$-basis is applied where $i,j$ run over unit cell positions $\mathbf{R}$. This non-atomistic description is referred to as effective bond-orbital model (EBOM) and is in line with the usage of a unit-cell averaged current operator for the calculation of transport properties as described in the next subsection.

The second and third term incorporate the modification due to the presence of randomly placed Mn impurities on the lattice: The variable $p_{i}$ takes the value $1$ with propability $x$ and $0$ with $1-x$, respectively, for each site $i$ in each different random configuration. While the on-site impurity potentials $V_{\alpha\alpha}$ are treated as free parameters to empirically adjust the energetic position of the impurity bands and $\hat{n}_{i\alpha\sigma}$ is the occupation operator, the local exchange constants $J_{\alpha d}$ couple impurity spins $\hat{\mathbf{S}}_{i}$ to carrier spins $\hat{\mathbf{s}}_{i,\alpha}$ of the system as described by the Zener $sp/d$-model. In particular, the $d$ electrons are assumed to be inner core electrons here and 
the use of a local contact interaction between the spin degrees of freedom corresponds to the most simple approach, see e.g. Refs. \onlinecite{benoit_a_la_guillaume_wigner-seitz_1992,dietl_hole_2008} for model extensions.

Due to the symmetry of the Ga-sublattice the on-site impurity potentials are anisotropic: $V_{p_{x}p_{x}}=V_{p_{y}p_{y}}\ne V_{p_{z}p_{z}}$. In accordance with the experiment, we assume the presence of a ferromagnetic solution for $T=0$ K and approximate the spin-spin interaction term with $\hat{\mathbf{S}}_{i}\approx\hat{S}_{i}^{z}\approx\langle\hat{S}_{i}^{z}\rangle=S$. This resembles a mean-field approximation by neglecting spin-flip scattering and using the fully spin-polarized ground state for $T=0$ K.

Furthermore, the off-diagonal hopping matrix elements for the Mn impurities (corresponding to inter-site impurity potentials) must be modified as well, as the simulations are carried out for a rather large concentration of impurities with $x=10\%$. This is achieved by introducing an additional scalar parameter $\gamma$ for the Mn-related hopping matrix elements which then tunes the bandwidth of possible Mn impurity bands. There are three different kinds of hopping matrix elements between GaN or MnN unit cells possible (the spin-index $\sigma$ has been suppressed for clarity):
\begin{eqnarray}
&t_{ij}^{\alpha\beta,\text{GaN-GaN}},&\\
&t_{ij}^{\alpha\beta,\text{MnN-MnN}}&=\gamma t_{ij}^{\alpha\beta,\text{GaN-GaN}},\\
&t_{ij}^{\alpha\beta,\text{GaN-MnN}}&=\frac{1+\gamma}{2}t_{ij}^{\alpha\beta,\text{GaN-GaN}}\label{tij_GaN_MnN}.
\label{eq:average_hopping}
\end{eqnarray}
At the GaN-MnN interface an arithmetically averaged hopping matrix element according to Eq.(\ref{eq:average_hopping}) is used to avoid the introduction of a further scaling parameter. The free parameters $V_{\alpha\alpha}$ and $\gamma$ were chosen such that the experimentally observed key features of the real part of the AC conductivity for photon energies $\hbar\omega>0.8$ eV were qualitatively reproduced. The values will be presented in the results section, where the long-wavelength regime is studied as a function of the Fermi level position and directly compared to the experimental data.

The value of $S=2$ is chosen according to the sample properties and we used $J_{sd}=0$\cite{suffczynski_effects_2011} in combination with an antiferromagnetic $J_{pd}=-1.2$ eV. The absolute values lie within the errorbars of Pacuski \textit{et al.}\cite{pacuski_excitonic_2007} and for a discussion of an apparent ferromagnetic sign we refer the reader to the theoretical work of Dietl. \cite{dietl_hole_2008}.

\subsection{Transport Properties}
The real part of the frequency-dependent conductivity (optical conductivity) is evaluated in linear response theory according to the standard expression for the diagonal matrix-elements of a noninteracting system:
\begin{eqnarray}
\sigma_{\xi\xi}^{\text{AC}}(\omega)&:=&\text{Re}\{\sigma_{\xi\xi}(\omega)\}\nonumber\\
&=&\frac{\pi}{V\omega}\sum_{n,m}\langle n\vert\hat{j}_{\xi}\vert m\rangle\langle m\vert\hat{j}_{\xi}\vert n\rangle\nonumber\\
&\times&\left[f(E_{n})-f(E_{m})\right]\delta(\omega+\omega_{n}-\omega_{m}).
\end{eqnarray}
Here, $\{\vert n\rangle\}$ is a complete set of eigenstates of the unperturbed Hamiltonian $\hat{H}$, $f(E)$ is the Fermi function, $V$ the system volume and $E_{n}=\hbar\omega_{n}$ are eigenenergies. The explicit expression for the current-operator $\hat{j}_{\xi}$ in our multiband EBOM is given by:
\begin{equation}
\hat{j}_{\xi}=-i\frac{e_{0}}{\hbar}\sum_{mn,\alpha\beta,\sigma}t_{mn}^{\alpha\beta,\sigma}(R_{\xi}^{n}-R_{\xi}^{m})\hat{c}_{m,\alpha,\sigma}^{\dagger}\hat{c}_{m,\beta,\sigma}.
\end{equation}
In this equation $i$ is the imaginary unit, $e_{0}$ the elementary charge, $m$ the electron mass, $\mathbf{R}^{m}$ a lattice site and $\xi$ denotes the corresponding vectorial component. This equation can be derived from the fundamental commutator relation for a single-particle Hamiltonian $\hat{H}$ and momentum $\hat{p}_{\xi}$,
\begin{equation}
\frac{i}{\hbar}[\hat{H},\hat{r}_{\xi}]_{-}=\frac{\hat{p}_{\xi}}{m},
\end{equation}
in combination with the envelope approximation for the spatial operator $\hat{\mathbf{r}}$:
\begin{equation}
\hat{\mathbf{r}}\approx\hat{\mathbf{R}}\approx\sum_{i,\alpha,\sigma}\mathbf{R}_{i}\hat{c}_{i,\alpha,\sigma}^{\dagger}\hat{c}_{i,\alpha,\sigma}.
\end{equation}
Consequently, microscopic contributions within the unit-cell are neglected in this current operator, see e.g. Ref. \onlinecite{tomczak_optical_2009} and references therein.

\subsection{Kernel polynomial method}
The numerical calculations in this paper were carried out using the kernel polynomial method (KPM)\cite{weisse_chebyshev_2004,weise_kernel_2006}. This approach allows for a direct calculation of the density of states of the system or traces over products of matrix elements of operators without exact diagonalization of the Hamiltonian $\hat{H}$.
The KPM uses a series expansion of $\hat{H}$ in Chebyshev polynomials $T_{n}(\hat{H})$ of order $n$. As described in detail in Refs. \onlinecite{weisse_chebyshev_2004,weise_kernel_2006}, the desired properties are then obtained from $\hat{H}$ by means of recursion relations when the expansion coefficients $\mu_{n}$ are systematically restricted up to a finite order. 
In this work, the Jackson kernel is used which corresponds to a Gaussian representation of the spectral densities' peaks. The broadening is analytically dependent on the order of the expansion. The choice of the order $n$ of the expansion depends on the purpose, e.g. resolving each spectral peak in the DOS as opposed to a smooth function.
In the present paper the more complicated case of a two-dimensional expansion is required for the evaluation of the real part of the frequency-dependent conductivity. According to the work of Weisse \textit{et al}\cite{weisse_chebyshev_2004,weise_kernel_2006}, the following function (which may be interpreted as a matrix element density) is defined,
\begin{equation}
j(E,E^{\prime})=\frac{1}{V}\sum_{n,m}\langle n\vert\hat{j}_{\xi}\vert m\rangle\langle m\vert\hat{j}_{\xi}\vert n\rangle\delta(E-\hbar\omega_{n})\delta(E^{\prime}-\hbar\omega_{m}),
\end{equation}
which can be directly calculated within the two-dimensional KPM. Then, the real part of the frequency-dependent conductivity is given by a double integration for various Fermi levels of interest:
\begin{eqnarray}
\sigma^{\text{AC}}_{\xi\xi}(\omega)&=&\frac{\pi}{\omega}\int_{-\infty}^{\infty}dE\int_{-\infty}^{\infty}dE^{\prime}j(E,E^{\prime})\nonumber\\
&\times& \left[f(E)-f(E^{\prime})\right]\delta(\hbar\omega-(E^{\prime}-E)).
\end{eqnarray}
The $\delta$-function is numerically approximated by a Lorentzian with uniform broadening, which is chosen to be approximately double the mean energy spacing in the conductivity calculations.

\subsection{Results}
\begin{figure}
\includegraphics[width=\columnwidth]{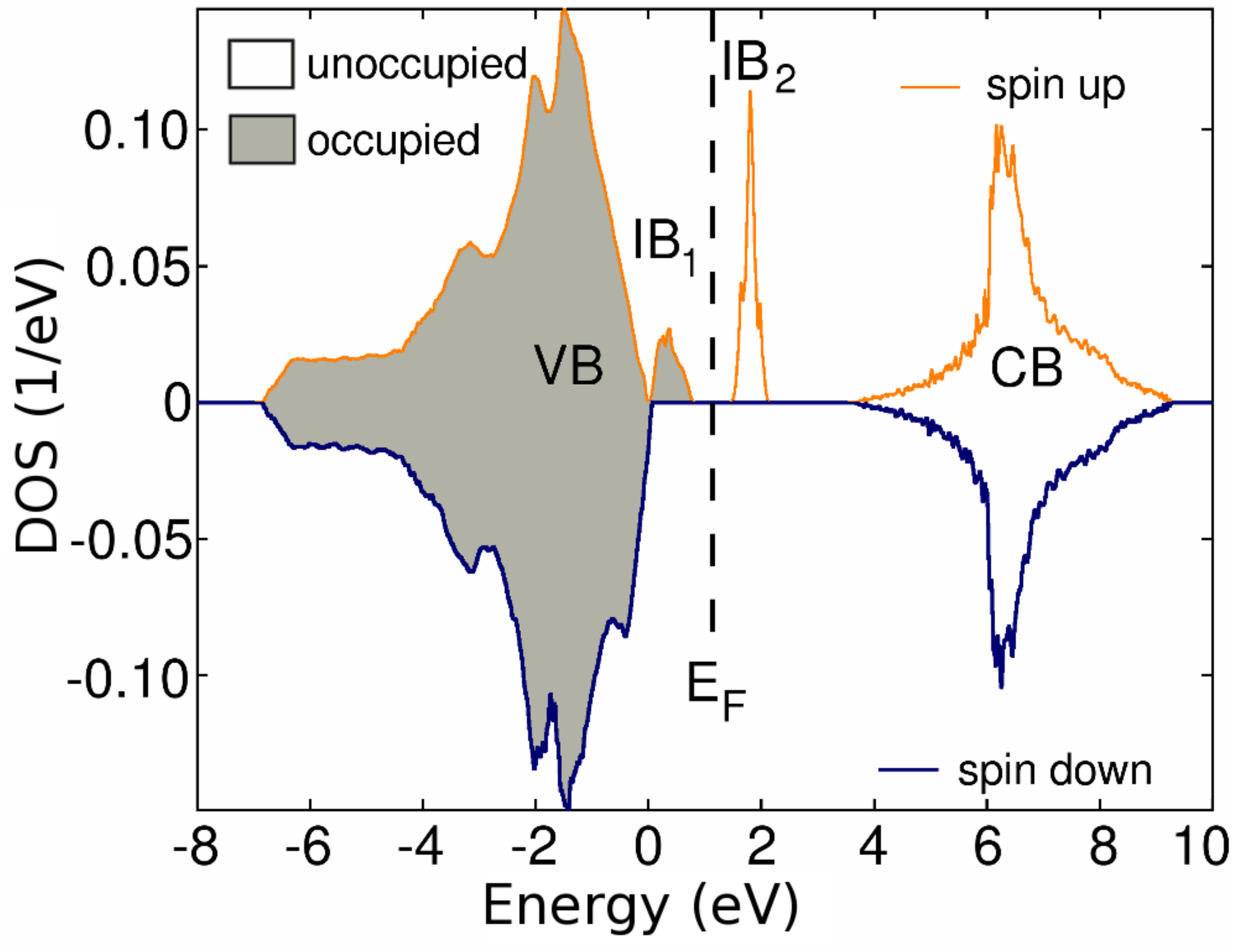}
\caption{\label{fig_2}Electronic density of states of Ga$_{1-x}$Mn$_{x}$N with $x=10\%$ determined within the effective-bond-orbital model for 0 K while a ferromagnetic solution was assumed. The valence band (VB) and lower impurity band (IB1) is filled, as the Fermi level lies in an impurity band gap below the upper impurity band (IB2) and the conduction band (CB).}
\end{figure}

When the free parameters of the electronic model are chosen to $V_{ss}=0$, $V_{p_{x}p_{x}}=V_{p_{y}p_{y}}=2.24$ eV and $V_{p_{z}p_{z}}=2.08$ eV including the scaling factor of $\gamma=0.35$ for the off-diagonal matrix elements, the characteristic features of the real part of the frequency-dependent conductivity for photon energies $\omega>0.8$ eV are qualitatively reproduced. Please mind that for reasons of mutual consistency the modeling was performed for a low temperature data set from Ref.\onlinecite{mourad_multiband_2010}, as the temperature dependence of the material parameters is not unambiguously known. The experimental measurement, on the other hand, were performed at room temperature. The calculations were carried out on supercells with $N=16$ unit cells in each spatial dimension (which results in a $8N^3\times8N^3$ matrix in total) with periodic boundary conditions. The disorder was simulated over 20 random configurations and we used 8192 moments in the KPM expansion for the DOS, which results in a satisfactorily smooth function. Fig.\ref{fig_2} shows the spin-resolved DOS. Here, one can observe two distinct impurity bands in the host band gap of GaN located at $\sim0.37$ eV and $\sim1.8$ eV. Between both impurity bands an impurity band gap of $\sim0.74$ eV exists. The spectral weight of these two bands are expected to have the ratio $1:2$ as $V_{p_{x}p_{x}}=V_{p_{y}p_{y}}\ne V_{p_{z}p_{z}}$ due to orbital symmetries of the $sp^3$-basis. We do find numerically a weight of $\sim0.0114$ and $\sim0.0250$ respectively, because some contributions are merged in the valence band. Our model DOS does not reproduce the spectral weight of 2:3 as caused by two non-bonding $e$-orbitals in relation to three non-bonding $t_{a}^{2}$-orbitals according to the single-particle picture given by Graf \textit{et al.}\cite{graf_mn3+/2+_2002,graf_prospects_2003} including $d$ orbitals. Nevertheless, the results are in good qualitative agreement with first-principle calculations\cite{kulatov_electronic_2002,uspenskii_ab_2003,titov_x-ray_2005,sato_first-principles_2010,virot_electronic_2011}.

\begin{figure}
\includegraphics[width=\columnwidth]{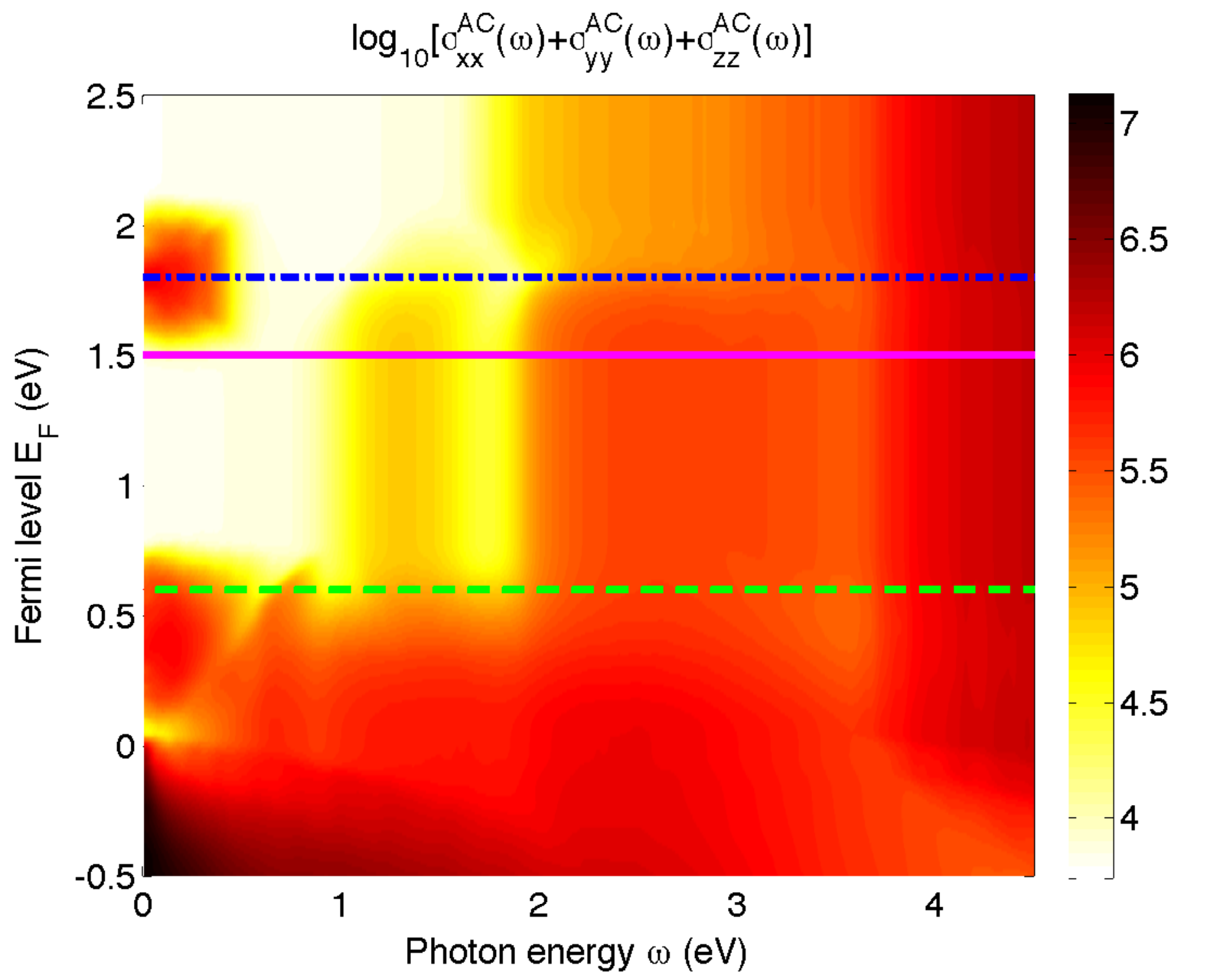}
\caption{\label{fig_3}Calculated real part of the AC-conductivity of Ga$_{0.9}$Mn$_{0.1}$N as a function of Fermi level and photon-energy relative to the valence band edge. The horizontal lines denote the corresponding intersections of the selected Fermi level positions visualized separately in Fig.\ref{fig_4}.}
\end{figure}

\begin{figure}
\includegraphics[width=\columnwidth]{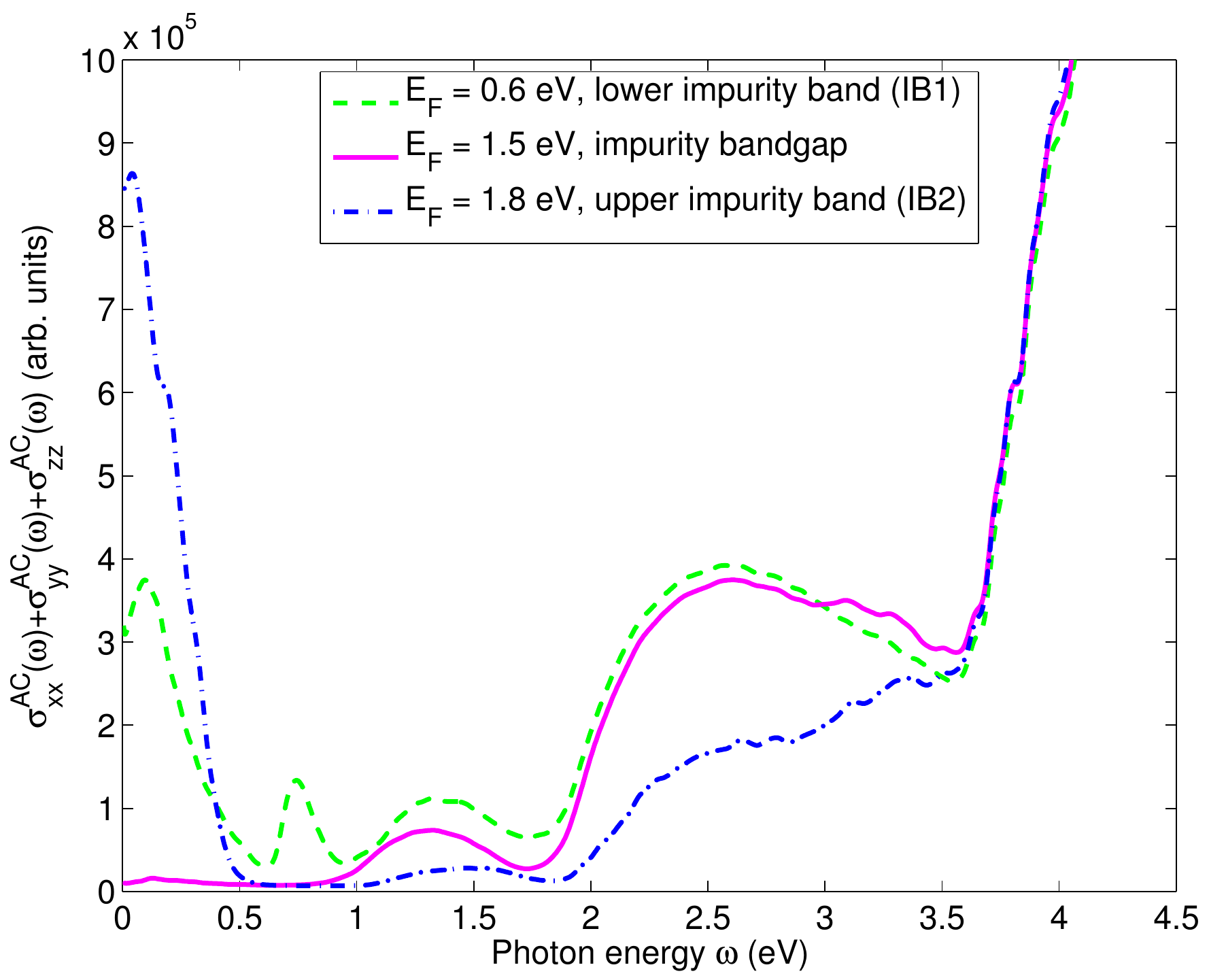}
\caption{\label{fig_4}Theoretically determined real part of the AC-conductivity of Ga$_{0.9}$Mn$_{0.1}$N for different Fermi level positions.}
\end{figure}

In Fig.\ref{fig_3} the trace over the diagonal matrix elements of the real part of the frequency-dependent conductivity is depicted in a logarithmic plot as a function of the photon energy $\omega$ and the Fermi level $E_{F}$. Additionally, we have visualized some manually selected curves for fixed Fermi levels in Fig.\ref{fig_4}.
The disorder average was taken over 10 random configurations and we used 2048 moments in the KPM expansion, as these calculations are numerically more demanding. We expect this choice to be sufficient for qualitative statements, but a finite-size analysis will be necessary if one is interested in extrapolating the zero-frequency limit\cite{weisse_chebyshev_2004}. This goes beyond the scope of the present study.

In Fig.\ref{fig_3} and Fig.\ref{fig_4}, we can easily identify a steep slope around 4 eV for positive Fermi energies $E_{F}$, the experimentally observed shoulder for $E_{F}\ge0.6$ eV (starting at $\omega\approx1.8$ eV) and the experimentally found peak around $\omega\approx1.3-1.4$ eV for $\sim0.6$ eV $\le E_{F}\le 1.9$ eV. Other values of the Fermi level lead to additional peaks or deviating peak positions. Therefore, we may conclude that the measured transport properties are in qualitative agreement with the theoretical calculations if the Fermi level falls either into the impurity band gap or lies in one of the two Mn-related impurity bands in the host band gap of GaN.

If we now turn the attention to the low-energy regime of $\omega\le0.8$ eV, the theoretical results predict the following features: If the Fermi level falls energetically into the upper impurity band with $\sim 1.6$ eV $\le E_{F}\le1.9$ eV, a peak structure for $\omega<0.5$ eV should appear. For a position in the lower impurity band with $E_{F}\approx0.6$ eV in addition to the former case, a second peak at about $\omega\approx0.75$ eV is predicted. In comparison to the ellipsometric measurements for $\omega\le0.8$ eV, which show the absence of any significant peak structure, we conclude that the Fermi level $E_{F}$ lies in the band gap between the two Mn-related impurity bands. This finding based on the present model study is in agreement with first-principle calculations of Virot \textit{et al.}\cite{virot_electronic_2011} which include electron correlations within LSDA+U and with experimental results for samples in the low concentration regime of $x\approx3.1\%$ that show highly resistive behaviour up to 300 K.\cite{bonanni_experimental_2011,sawicki_origin_2012}

As the used EBOM is discretized on the hexagonal Bravais lattice in contrast to the atomistic wurtzite-structure, a calculation of the Heisenberg exchange integrals (not shown) and solution of the disordered Heisenberg model in Tyablikow decoupling (not shown) was performed with the $C_{6v}$ symmetry. A ferromagnetic solution was found with $T_C\approx15$ K in the case of the Fermi level lying in the upper impurity band, i.e. corresponding to the half-metallic scenario observed in LSDA\cite{virot_electronic_2011}. In contrast, for a Fermi level lying in the impurity band gap, no ferromagentic solution could be found as the effective exchange integrals were mainly antiferromagnetic. Therefore we conclude, that the \textit{sp/d}-exchange mechanism used within this particular multiband model cannot explain both transport and magnetic properties at the same time. An extension to a fully atomistic model is beyond the scope of this work, as the then obtained 16 bands per spin direction would not allow for a reasonable disorder average due to current computational limits. Furthermore, the simulation of the impurity band position would require more free parameters.

\section{Summary}
In this paper, wurtzite Ga$_{1-x}$Mn$_{x}$N epitaxial films deposited by MBE with $x=4\%$ \& $x=10\%$ have been investigated by means of ellipsometry for wavelengths between 200--12400 nm. The characteristic features of the real part of the frequency-dependent conductivity known from samples with less Mn content produced by MOVPE\cite{bonanni_experimental_2011,sawicki_origin_2012} could be confirmed for the layers\cite{kunert_ga1xmnxn_2012} with high Mn-content studied in the present paper. In particular, a steep slope in the GaN host band gap region, a shoulder starting at about 2 eV and the characteristic peak structures at 1.31 eV ($x_{\text{Mn}}=10\%$) and 1.35 eV ($x_{\text{Mn}}=4\%$) are observed experimentally. To understand the relatively low Curie temperatures and exchange coupling mechanism in this material, the knowledge of the electronic structure and Fermi level position is crucial. Both were determined by model calculations using the multiband effective-bond-orbital model with a $sp^3$-basis combined with the V-J model to describe the Mn-impurities including disorder effects. The diagonal and off-diagonal hopping matrix elements are adjusted in order to qualitatively reproduce the characteristic features of the real part of the frequency-dependent conductivity for photon energies $\omega>0.8$ eV. The corresponding calculations are carried out within linear response theory using the kernel polynomial method.

Our model calculations indicate the existence of two separated Mn-related impurity bands lying at 0.37 eV and 1.8 eV in the host band gap of GaN. If the Fermi level falls into one of those two impurity bands, significant peak structures for photon energies $\omega<0.8$ eV are theoretically predicted in the real part of the frequency-dependent conductivity. Their experimental absence indicates a Fermi level position in the impurity band gap of about $0.74$ eV, which strongly indicates resistive behaviour for large wavelengths. Even for these rather large Mn concentrations, the samples appear to be insulating. The fact that the material is ferromagnetic indicates that the model as used in this paper alone (in combination with our approximations) is probably not sufficient to describe the transport properties and the magnetism simultaneously. Therefore, an improved theoretical model is necessary to calculate the Heisenberg exchange integrals and the magnetic properties. Nevertheless, these results help to understand the origin of the insulating behaviour of the material. Furthermore, sufficient copoding might lead to a shift of the Fermi level into a region of finite density of states and thus enable direct exchange mechanisms.

\section{Acknowledgements}
We gratefully acknowledge the financial support from the DFG - 436 RUM 113/27/0-2 as well as the support by the European Community through the FunDMS Advanced Grant (No. 227690) of the ERC within the `Ideas' 7th Framework Programme. The authors would like to thank Thomas Wagner (Lot-QuantumDesign GmbH, Germany), Tom Tiwald (J.A. Woollam Co., Inc, USA.) and Tomasz Dietl for their support and fruitful discussions. Stefan Barthel thanks Georges Bouzerar for an introduction into the field of DMS.

\end{document}